# On the Equilibrium of Query Reformulation and Document Retrieval

Shihao Zou
University College London
shihao.zou.17@ucl.ac.uk

Guanyu Tao
University College London
g.tao@ucl.ac.uk

Jun Wang
University College London
junwang@cs.ucl.ac.uk

Weinan Zhang
Shanghai Jiao Tong University
wnzhang@sjtu.edu.cn

Dell Zhang
Birbeck, University of London
dell.z@ieee.org

## ABSTRACT

In this paper, we study jointly query reformulation and document relevance estimation, the two essential aspects of information retrieval (IR). Their interactions are modelled as a two-player *strategic game*: one player, a query formulator, taking actions to produce the optimal query, is expected to maximize its own utility with respect to the relevance estimation of documents produced by the other player, a retrieval modeler; *simultaneously*, the retrieval modeler, taking actions to produce the document relevance scores, needs to optimize its likelihood from the training data with respect to the refined query produced by the query formulator. Their *equilibrium* or *equilibria* will be reached when both are the best responses to each other. We derive our equilibrium theory of IR using normal-form representations: when a standard relevance feedback algorithm is coupled with a retrieval model, they would share the same objective function and thus form a *partnership game*; by contrast, pseudo relevance feedback pursues a rather different objective than that of retrieval models, therefore the interaction between them would lead to a *general-sum game* (though implicitly collaborative). Our game-theoretical analyses not only yield useful insights into the two major aspects of IR, but also offer new practical algorithms for achieving the equilibrium state of retrieval which have been shown to bring consistent performance improvements in both text retrieval and item recommendation.

## CCS CONCEPTS

• **Information systems** → **Query reformulation**; **Collaborative search**; *Learning to rank*; *Relevance assessment*;

## KEYWORDS

relevance feedback; retrieval model; game theory

## 1 INTRODUCTION

In information retrieval (IR), we have two distinctive yet correlated research challenges. The first challenge lies in how to formulate optimal queries in order to best represent the user's information needs. In text retrieval, a fundamental theory is Rocchio's relevance feedback [26], i.e., in a vector space model setting, the optimal query reformulation is achieved by making use of relevance feedback [7]. The idea is to have an iterative process for a search system where the system takes the results that are initially returned from a given query, gathers user feedback, and then utilizes the information about whether or not those results are relevant to refine and expand the terms in the query. In practice, the feedback signal could be implicit (such as clicks or playlists) or blinded (assuming the first top-$k$ returned as relevant ones), depending on the specific cases [30]. In the case that no initial query is given such as in collaborative filtering based recommender systems, the user's information need completely relies on the "relevance feedback" from historical interactions and may be inferred from other similar user profiles [28]. The second challenge is concerned with relevance estimation where the fundamental goal is to assign a relevance score for each of the documents given the information need representation (i.e., query). The classic information retrieval model [25], its extension BM25 [23], and the language models [34] all utilize term weighting in order to devise the document relevance scores, whereas the latest machine learning methods such as learning to rank [21] and the generative adversarial net (GAN) approach [29] make use of historical relevance judgements to directly learn the document scores.

In this paper, we aim to bring the two challenges together in a single unified framework. Our study is based on game-theoretical analysis. We assume that the information need does not change throughout the session. The new equilibrium theory of information retrieval states that, instead of considering the above two challenges separately, there is a *strategic game* played *simultaneously* between the query reformulation model and the retrieval model. More specifically, the query reformulation player would refine the query that is *the best response* to the actions from the given retrieval model player, i.e., formulate an optimal query that would maximize its utility (e.g., the score difference between relevant documents and non-relevant documents) given the retrieval model. At the same time, the retrieval model player would also need to produce the document relevant estimation that is *the best response* toward the formulated query, i.e., the relevance estimation would maximize the retrieval utility (e.g., the likelihood function from the historical user relevance judgements), given the query.

Thus, the game play provides the retrieval solution(s) that is in a *Nash equilibrium* or multiple *Nash equilibria* [20] when each becomes the best response to the other. This game-theoretical information retrieval solution is, however, a general one, as different definitions of query reformulation and relevance estimation utilities would lead to different ways approaching the retrieval process equilibrium or equilibria. In this paper, we show that a joint play between the standard relevance feedback and the document retrieval model would lead to a *partnership game* [27], whereas the pseudo relevance feedback based query expansion would result in a *general-sum game* [20] with the retrieval model. A practical implementation is performed for evaluating our game-theoretical approach in both text retrieval and collaborative filtering tasks. Our results show that an equilibrium solution to relevance feedback consistently outperforms other methods modelling query reformulation and relevance estimation separately. Just using a linear retrieval model our framework would improve the overall retrieval performance in the case of pseudo relevance feedback. Furthermore, compared with the simpler case where both components have the same utility function, our framework provides better results with a higher convergence rate.

## 2 RELATED WORK

In information retrieval, Rocchio performed the first work [26] to investigate how to incorporate relevance feedback for refining queries in a search system. The main idea behind Rocchio's algorithm is to update the initial query according to the user's feedback that whether a document in the collection is relevant or not. The goal for Rocchio's query reformulation is to identify an optimal query that maximizes the separation between relevant documents and non-relevant documents from the scores produced by a particular retrieval model, namely

$$\mathbf{u}_R = \frac{1}{|D_r|} \sum_{\mathbf{d}_i \in D_r} \rho(\mathbf{d}_i, \mathbf{q}) - \frac{1}{|D_n|} \sum_{\mathbf{d}_i \in D_n} \rho(\mathbf{d}_i, \mathbf{q}), \quad (1)$$

where $\mathbf{q}$ is a query vector output from the query formulator and $D_r$ and $D_n$ are the sets of relevant and non-relevant document vectors denoted by $\mathbf{d}$. $\rho(\mathbf{d}_i, \mathbf{q})$ is the retrieval model, producing the relevance score for a document $\mathbf{d}_i$ given a query $\mathbf{q}$. In essence, optimizing the objective would allow terms to be selected from relevant documents and the initial query to be expanded when relevance feedback is given [10, 12]. Apart from relevance feedback based query expansion, query can also be expanded via external knowledge resources such as dictionaries, WordNet [19], ConceptNet [15] and Wikipedia [14].

Pseudo relevance feedback (or retrieval feedback) [32], however, is a more practical approach as users are usually not involved in the process of relevance feedback and retrieved results are the only feedback. In this case, Rocchio's algorithm can be modified for pseudo relevance feedback if assuming top-$k$ retrieved documents are relevant while others are non-relevant. Query expansion and reformulation have also been studied in the framework of Binary Independence Model [25] including Robertson Selection Value (RSV) [24] and Kullback-Leibler divergence (KLD) [4]. Recently, PMING distance [6] and word embedding technique [18] have also been widely applied in IR. Besides term features, feedback models are also investigated. A term classification process [3] and a posterior distribution for the feedback model [5] are proposed. A cluster-based resampling method is presented in [13].

In this paper we study query reformulation but in a more general setting where the learning of retrieval model is simultaneously considered. Our work departs from that of [11, 25] whose focus is on the term weighting only. Our work is also significantly different from the risk minimization framework [11] in that we recognize the fact that the objectives from the query process and the retrieval process, although correlated, may not be necessarily identical (particularly in pseudo relevance feedback as we will illustrate later). Interactive interface is studied in [17, 33, 35] where a search session is modeled as a dual-agent stochastic game in [17] and card playing in [35]. Our work is different from them in that we also investigate different objectives for the query and retrieval model as a general-sum game.

Modeling the interactions between them as a strategic game would provide a sound foundation for understanding their discrepancies such as query drifts and its stability issue [5] (more discussion will be provided in the next section). This paper is built upon the foundation of game theory [20] in particular normal-form game. In a strategic game, each player has a reward that depends on the strategies of all the players rather than itself only. In order to achieve its goal, the player learns a policy to perform actions based on the observations of other players' strategies and the environment. In a Nash equilibrium each player's choice is the best response to the others and no player can gain by unilateral deviation from the joint strategy. The above provides a sound theoretical foundation for us to understand the retrieval processes which may have different objectives. Under the game-theoretical framework, we unify the relevance feedback and retrieval modeling, and the two components are expected to be able to help each other and finally improve the performance of retrieval.

## 3 THE EQUILIBRIUM THEORY OF IR

In this section, we start with the problem settings, and then present our IR game theory with relevance feedback and pseudo relevance feedback and discuss their differences.

In an IR system, the user submits an initial query which (roughly) represents the information need. The query is denoted by a vector $\mathbf{q}$ with the dimension of vocabulary size, where its element $\mathbf{q}_j$ represents the binary occurrence of a term $j$ in the query. Each document $\mathbf{d}_i$ is represented by a binary vector over the term vocabulary $V$, where each dimension $\mathbf{d}_j$ represents whether the term $j$ is contained in the document (1) or not (0). We focus on binary relevance. Specifically, given the user's information need clued by the query, each candidate document $\mathbf{d}_i$ has an underlying ground-truth relevance score $r_i$, which is 1 for relevant documents and 0 for non-relevant ones. The IR task is to find relevant documents given a new query with 1) the information about previous queries and their relevance judgements, and 2) possibly some initial relevant judgements for the new query. To achieve the task, one can build up a refined query and a retrieval model to assess the relevance of the documents in the collection.

In this paper, the IR task is modelled in a normal-form game [20]. Formally, the game has three elements: two players (query formulator $Q$ and retrieval model $M$), the pure strategies $\{S_Q, S_M\}$ for player $Q$ and $M$ and the utility functions $\{u_Q, u_M\}$.

*Definition 3.1.* An **IR Strategic Game** is a tuple $\langle P, S, U \rangle$, where
- $P = \{Q, M\}$ is the set of two players: query formulator $Q$ and retrieval model $M$;
- $S = S_Q \times S_M$, where $S_Q$ and $S_M$ are finite sets of strategies available to player $Q$ and $M$. The strategies of the two players are denoted by $\mathbf{s}_q$ and $\mathbf{s}_m$ respectively where $\mathbf{s}_q \in S_Q$, $\mathbf{s}_m \in S_M$. In text retrieval, $\mathbf{s}_q$, for instance, is a binary vector of vocabulary size $|V|$ with its element representing whether the term is included in the query or not. On the other hand, $\mathbf{s}_m$ is a *weight* vector of the retrieval model that instantiates a retrieval model to distinguish the relevant documents from the non-relevant ones given the query $\mathbf{q}$.
- Finally, an equilibrium state should be achieved when both of them have no incentive to change their strategies $\mathbf{s}_q^*$ and $\mathbf{s}_m^*$, where $u_Q(\mathbf{s}_q^*, \mathbf{s}_m^*) \geq u_Q(\mathbf{s}_q, \mathbf{s}_m^*)$ and $u_M(\mathbf{s}_q^*, \mathbf{s}_m^*) \geq u_M(\mathbf{s}_q^*, \mathbf{s}_m)$ for any $\mathbf{s}_q$ and $\mathbf{s}_m$.

In practice, we consider a mixed strategy case where each strategy of the player is assigned a probability, and we assume that each element (such as a term, a document or an item) in a strategy is independent of each other. On the top of it, we can approximate the probability by the likelihood of all the elements in the strategy. The details are discussed below.

### 3.1 IR Game with Relevance Feedback

For relevance feedback task, the utilities of query formulator $Q$ and retrieval model $M$ are the same, i.e., the degree of successfully distinguishing relevant documents from the non-relevant ones. The common utility $u(\mathbf{s}_q, \mathbf{s}_m)$ can be defined as

$$u(\mathbf{s}_q, \mathbf{s}_m) = \frac{1}{|D_r|} \sum_{\mathbf{d}_i \in D_r} \log p(r = 1 | \mathbf{d}_i, \mathbf{q}; \theta) - \frac{1}{|D_n|} \sum_{\mathbf{d}_i \in D_n} \log p(r = 0 | \mathbf{d}_i, \mathbf{q}; \theta), \quad (2)$$

where $D_r$ and $D_n$ are the ground-truth of relevant and non-relevant document sets for the user's information need respectively and $\theta$ represents the relevance score given by the retrieval model. In such a case, the IR game becomes a *partnership game* [27], where each player receives an identical reward and acts independently while optimizing the above global objective function. The Nash equilibria correspond to local optima of the objective function. It is worth mentioning that a partnership game allows each player to have a different cost function, which in our case would be different constraints or learning regularizations reinforced for the query reformulation method and the retrieval model, independently and separately.

To illustrate this, we present a simple example of the IR game with relevance feedback. Suppose we have two terms $t_1$ and $t_2$ and two documents $\mathbf{d}_1$ and $\mathbf{d}_2$. One can depend on Binary Independence

**Table 1: An IR game example (relevance feedback).**

|   | $\mathbf{d}_1$ | $\mathbf{d}_2$ |
|---|---|---|
| $t_1$ | 1 | 0 |
| $t_2$ | 0 | 1 |
| $r$ | 1 | 0 |

(a) Corpus

|   | $\mathbf{s}_{m_1} =$ $\{1, 0.2\}$ | $\mathbf{s}_{m_2} =$ $\{0.2, 1\}$ |
|---|---|---|
| $\mathbf{s}_{q_1} = \{1, 0\}$ | -1.0064 | -1.2913 |
| $\mathbf{s}_{q_2} = \{0, 1\}$ | -1.4913 | -2.0064 |

(b) Utilities of Strategies

Model [25] to establish each term's underlying correlation with the document relevance, given the information need (no need a specific query).

Table 1a gives example term binary occurrence in the documents and the underlying relevance of each term, where $D_r = \{\mathbf{d}_1\}$ and $D_n = \{\mathbf{d}_2\}$. Table 1b shows the corresponding common utilities w.r.t. different strategies of the two players. $\mathbf{s}_{q_1} = \{1, 0\}$ means the query is $t_1$ only and $\mathbf{s}_{q_2} = \{0, 1\}$ means the query is $t_2$ only. Suppose the model parameter space has only two points: $\mathbf{s}_{m_1} = \{1, 0.2\}$ means to assign the weight 1 to $\theta_1$ for $\mathbf{d}_1$ and 0.2 to $\theta_2$ for $\mathbf{d}_2$; $\mathbf{s}_{m_2} = \{0.2, 1\}$ means to assign the weight 0.2 to $\theta_1$ and 1 to $\theta_2$ to calculate the query-document relevance score $p(r = 1 | \mathbf{d}_i, \mathbf{q}; \theta)$ as

$$p(r = 1 | \mathbf{d}_i, \mathbf{q}; \theta) = \text{sigmoid}\left(\theta_1 \mathbf{q}_1 \mathbf{d}_{i1} + \theta_2 \mathbf{q}_2 \mathbf{d}_{i2}\right) \quad (3)$$

and based on which ranks them and selects the top one. The utility score is based on the underlying relevance score of the returned document by the retrieval model.

It is easy to calculate that this IR game has a Nash equilibrium at $(\mathbf{s}_{q1} = \{1, 0\}, \mathbf{s}_{m1} = \{1, 0.2\})$. In such a case, via Eq. (3) we have

$$p(r = 1 | \mathbf{d}_1, \mathbf{q}; \theta) = \text{sigmoid}(1 \times 1 \times 1 + 0.2 \times 0 \times 0) = 0.7311$$
$$p(r = 1 | \mathbf{d}_2, \mathbf{q}; \theta) = \text{sigmoid}(1 \times 1 \times 0 + 0.2 \times 0 \times 1) = 0.5$$

and thus via Section 3.1 the common utility is

$$u(\mathbf{s}_q, \mathbf{s}_m) = \log p(r = 1 | \mathbf{d}_1, \mathbf{q}; \theta) + \log p(r = 0 | \mathbf{d}_2, \mathbf{q}; \theta) = -1.0064.$$

With the iteration between the two players, the updated query and retrieval model will both converge to the equilibrium, which in a more practical setting corresponding to the local optimal of the objective.

### 3.2 IR Game with Pseudo Relevance Feedback

For pseudo relevance feedback task, the utilities of query formulator $Q$ and retrieval model $M$ are different. Specifically, the utility of the retrieval model is still the degree of successfully distinguishing relevant documents from the non-relevant ones, which we copy here:

$$u_M(\mathbf{s}_q, \mathbf{s}_m) = \frac{1}{|D_r|} \sum_{\mathbf{d}_i \in D_r} \log p(r = 1 | \mathbf{d}_i, \mathbf{q}; \theta) - \frac{1}{|D_n|} \sum_{\mathbf{d}_i \in D_n} \log p(r = 0 | \mathbf{d}_i, \mathbf{q}; \theta). \quad (4)$$

However, the utility of the query formulator $Q$ is solely based on the feedback from the retrieval model $M$

$$u_Q(\mathbf{s}_q, \mathbf{s}_m) = \frac{1}{|D_k|} \sum_{\mathbf{d}_i \in D_k} \log p(r = 1 | \mathbf{d}_i, \mathbf{q}; \theta) - \frac{1}{N - |D_k|} \sum_{\mathbf{d}_i \notin D_k} \log p(r = 0 | \mathbf{d}_i, \mathbf{q}; \theta), \quad (5)$$

Table 2: An IR game example (pseudo relevance feedback).

(a) Corpus

|   | $d_1$ | $d_2$ |
|---|---|---|
| $t_1$ | 1 | 0 |
| $t_2$ | 0 | 1 |
| $r$ | 1 | 0 |

(b) Utilities of Strategies ($u_Q$, $u_M$)

|   | $s_{m_1} = \{1, 0.2\}$ | $s_{m_2} = \{0.2, 1\}$ |
|---|---|---|
| $s_{q_1} = \{1, 0\}$ | (-1.0064, -1.0064) | (-1.2913, -1.2913) |
| $s_{q_2} = \{0, 1\}$ | (-1.2913, -1.4913) | (-1.0064, -2.0064) |

where $D_k$ is the top-$k$ document list ranked by the retrieval model given the query representation **q** and $N$ is the number of documents in the dataset. The discrepancies between Sections 3.2 and 3.2 indicate that the utilities of two players are not necessary the same in pseudo relevance feedback scenarios and the IR retrieval game only represents an implicit coordination between them.

To illustrate this, we present a simple toy example for the pseudo relevance feedback setting. Similar to Table 1a, here Table 2a presents the sample term binary occurrence in the documents and the underlying relevance of each term. Table 2b shows the corresponding utilities w.r.t. different strategies of the two players. The pure strategies of two players are set as the same as in Table 1b. And the query-document relevance score $p(r = 1|\mathbf{d}_i, \mathbf{q}; \theta)$ is calculated just as in Eq. (3). The utility scores $u_q$ and $u_m$ are then calculated by Sections 3.2 and 3.2 respectively. For example, when ($\mathbf{s}_{q2} = \{0, 1\}, \mathbf{s}_{m1} = \{1, 0.2\}$), we have

$p(r = 1|\mathbf{d}_1, \mathbf{q}; \theta) = \text{sigmoid}(1 \times 0 \times 1 + 0.2 \times 1 \times 0) = 0.5$
$p(r = 1|\mathbf{d}_2, \mathbf{q}; \theta) = \text{sigmoid}(1 \times 0 \times 0 + 0.2 \times 1 \times 1) = 0.5498.$

Thus $\mathbf{d}_2$ ranks higher than $\mathbf{d}_1$, i.e., $\mathbf{d}_2 \in D_k$ and $\mathbf{d}_1 \notin D_k$ here. As such, the utility of $Q$ and $M$ are

$u_Q(\mathbf{s}_{q_2}, \mathbf{s}_{m_1}) = \log p(r = 1|\mathbf{d}_2, \mathbf{q}; \theta) + \log p(r = 0|\mathbf{d}_1, \mathbf{q}; \theta) = -1.2913$
$u_M(\mathbf{s}_{q_2}, \mathbf{s}_{m_1}) = \log p(r = 1|\mathbf{d}_1, \mathbf{q}; \theta) + \log p(r = 0|\mathbf{d}_2, \mathbf{q}; \theta) = -1.4913,$

which are different because the ranking is not the same as the ground-truth.

From Table 2b we can observe that (i) $u_Q$ and $u_M$ can largely depart when the query representation is at a wrong region (e.g., $\mathbf{s}_{q_2}$). (ii) When the query representation is already at the wrong region (e.g., $\mathbf{s}_{q_2}$), the pseudo relevance feedback model could result in a wrongly optimistic assessment (e.g., $u_Q(\mathbf{s}_{q_2}, \mathbf{s}_{m_2}) > u_M(\mathbf{s}_{q_2}, \mathbf{s}_{m_2})$) (see the down-right corner). This explains the reason why pseudo relevance feedback would easily result in query drifts if no constraint was reinforced. With pseudo relevance feedback alone, it is difficult to spot the drifts given the high utility value of the query reformulation. However, as shown in the down-right corner, the utility of the retrieval model is in fact very low, which provides a strong indication of the drifts and potentially prevents them when strategically considering the retrieval model in the game. (iii) Nonetheless, there is still a Nash equilibrium at ($\mathbf{s}_{q_1} = \{1, 0\}, \mathbf{s}_{m_1} = \{1, 0.2\}$), which is the unique optima of this IR game.

## 4 APPLICATIONS

We study the applications of IR game on two typical IR scenarios, namely ad-hoc text retrieval and collaborative filtering based recommender systems.

### 4.1 Text Retrieval

In the text retrieval task, in order to find a good representation of information need, the query **q** tries to assign higher probability to terms only existing in relevant documents instead of those in non-relevant documents. On the other hand, the retrieval model tries to make the prediction more accurate given the current query information representation. For concise notation, we use $\theta_i$ to denote the predicted relevance between **q** and $\mathbf{d}_i$ by the retrieval model and $\phi$ indicates the parameter of a linear retrieval model which we will use later. We analyze several different equilibrium learning cases that are possible to happen in the game for text retrieval. Our algorithm is closely related to *Fictitious Play* [1] that each player always plays the best response of others at each iteration using gradient ascent.

*4.1.1 Case 1: Query Iteration.* In this case, the parameter $\phi$ of the retrieval model $M$ is fixed and the query formulator $Q$ is updated iteratively via Gradient Ascent. We use vector inner product to calculate the predictions

$$\theta_i = \text{sigmoid}(\mathbf{q}^\top \mathbf{d}_i) = \frac{1}{1 + e^{-\mathbf{q}^\top \mathbf{d}_i}}. \quad (6)$$

Thus $Q$ could constantly update **q** by maximizing $u_Q(\mathbf{s}_q, \mathbf{s}_m)$ as in Section 3.2 with gradient methods:

$$\frac{\partial u_Q(\mathbf{s}_q, \mathbf{s}_m)}{\partial \mathbf{q}} = \frac{1}{|D_r|} \sum_{d_i \in D_r} (1 - \theta_i)\mathbf{d}_i - \frac{1}{|D_n|} \sum_{d_i \in D_n} \theta_i \mathbf{d}_i, \quad (7)$$

$$\mathbf{q} \leftarrow \mathbf{q} + \eta \frac{\partial u_Q(\mathbf{s}_q, \mathbf{s}_m)}{\partial \mathbf{q}}, \quad (8)$$

where $\eta$ is the learning rate.

If there is only one retrieval process and no feedback provided, we call it *naive retrieval*. Besides, if the game has only one round where relevance feedback is produced only once, the case reduces to *Rocchio*'s algorithm [26]. If infinite rounds are allowed in this game, the query reformulation is expected to converge to a state where its utility $u_Q$ is maximized. We name it *Conv-Q*.

*4.1.2 Case 2: Retrieval Model Iteration.* In this case, the query representation is fixed while the retrieval model tries to maximize its utility as shown in Section 3.2. For generality, we define a simple linear retrieval model whose relevance score is the weighted sum of three different weighting schemes (result fusion), including binary vector space model (VSM), TDIDF and BM25. The document $\mathbf{d}_i$ in the $k$-th weighting scheme can be represented as a vector $\mathbf{d}_i^k$ of vocabulary size and the query has the same representation $\mathbf{q}^k$. Thus the representation of document $\mathbf{d}_i$ and the query **q** is a set of feature vectors $\mathbf{d} = \{\mathbf{d}^1, ..., \mathbf{d}^K\}$ and $\mathbf{q} = \{\mathbf{q}^1, ..., \mathbf{q}^K\}$, where $K = 3$. Cosine distance is used to determine the relevance between a document and the query in each weighting scheme. The score $\theta_i$ assigned to document $\mathbf{d}_i$ is the weighted sum of scores in all $K$ weighting schemes:

$$\theta_i = \text{sigmoid}\left(\sum_{k=1}^{K} w_k \cdot (\mathbf{d}_i^k)^\top \mathbf{q}^k\right), \quad (9)$$

where $\phi = \{w_1, ..., w_K\}$ are parameters of the linear retrieval model. We can update $w_k$ according to

$$\frac{\partial u_M(\mathbf{s}_q, \mathbf{s}_m)}{\partial w_k} = \frac{1}{|D_r|} \sum_{\mathbf{d}_i \in D_r} (1 - \theta_i) \cdot (\mathbf{d}_i^k)^\top \mathbf{q}^k - \frac{1}{|D_n|} \sum_{\mathbf{d}_i \in D_n} \theta_i \cdot (\mathbf{d}_i^k)^\top \mathbf{q}^k, \quad (10)$$

$$w_k \leftarrow w_k + \eta \frac{\partial u_M(\mathbf{s}_q, \mathbf{s}_m)}{\partial w_k}, \quad (11)$$

to maximize the utility. Finally, the retrieval model is expected to converge to an optima where the retrieval model utility $u_M$ is maximized. We name such a case as *Conv-M*.

In fact, there are many other learnable retrieval models such as support vector machines, boosted trees and neural networks [2]. Since our focus in this paper is not the retrieval model itself, but the IR game theory, we will not explore these sophisticated models here.

*4.1.3 Case 3: Equilibrium of the Query and Retrieval Model.* In this case, both the query formulator $Q$ and the retrieval model $M$ try to maximize their own utility. But at the same time, their strategies are involved with each other in each round of the game. The strategy of the retrieval model depends on the information representation of the query, and in turn, the strategy of the query formulator depends on the retrieval results to adjust the weight of each term in the query. Therefore, in each round of the game, the query and the retrieval model can be updated as Sections 4.1.1 and 4.1.2. Finally, both of them are expected to reach an equilibrium where the query and the retrieval model would not change their strategy. We call this case *Equil-Q&M*.

## 4.2 Item Recommendation

In this section, we extend our solution to personalized recommendation scenario, where there is no explicit query terms but the user profiles are represented by the items that have been consumed historically from the users.

*4.2.1 User-based item recommendation.* User-based recommendation is quite similar to text retrieval, where the vocabulary can be regarded as a pool of items and the initial query is the profile of the target user. To be specific, the target user's initial profile is the set of items that were rated previously. We use vector $\mathbf{q}_u$ to represent the target user's profile. Each element in $\mathbf{q}_u$ represents the rating given by the user to the item and 0 otherwise. In the memory based, there is a set of $N$ users in total whose profiles can also be represented as a vector, $\{\mathbf{d}_1, ..., \mathbf{d}_i, ..., \mathbf{d}_N\}$. On the top of it, the query is equivalent to the target user's profile $\mathbf{q}_u$, and the retrieval model is to predict the relevance between the target user and any other user, similarly denoted by $\theta_i = p(r = 1|\mathbf{d}_i, \mathbf{q}_u)$. However, in the user-based recommendation task, it is difficult to directly obtain the relevance $r$ between target user and other users. Thus only pseudo relevance feedback can be applied, in which the top-$k$ similar users are considered to be relevant neighbors while the others are non-relevant. $D_r$ and $D_n$ denote the sets of relevant and non-relevant users respectively.

There are three cases in the user-based item recommendation, similar to the cases discussed in Section 4.1. The first case is *Rocchio*'s algorithm and *Conv-Q* where only the target user's profile is updated according to Section 4.1.1. In the second case (*Conv-M*), we define the retrieval model as

$$\theta_i = p(r = 1|\mathbf{d}_i, \mathbf{q}_u) = \text{sigmoid}(\mathbf{q}_u^\top \mathbf{W} \mathbf{d}_i), \quad (12)$$

where $\mathbf{W}$ is the parameters of the retrieval model. Using gradient descent, $\mathbf{W}$ can be updated by

$$\mathbf{W} \leftarrow \mathbf{W} + \eta \left( \frac{1}{|D_r|} \sum_{\mathbf{d}_i \in D_r} (1 - \theta_i) \mathbf{q}_u \mathbf{d}_i^\top - \frac{1}{|D_n|} \sum_{\mathbf{d}_i \in D_n} \theta_i \mathbf{q}_u \mathbf{d}_i^\top \right). \quad (13)$$

The third case (*Equil-Q&M*) is that both target user's profile $\mathbf{q}_u$ and retrieval model parameter $\mathbf{W}$ are updated in each round of the game according to Section 4.1.1 and Eq. (13) for different utility functions, respectively, using gradient ascent.

## 5 EXPERIMENTS

Using public benchmarking datasets, we have conducted preliminary experiments on both text retrieval and item recommendation, in order to verify our proposed theory of IR game equilibrium.

### 5.1 Text Retrieval

In the text retrieval task, we use the set of documents in TREC disks 4 & 5 [8]. There are 250 topics along with a collection of judged documents. The relevance judgements are given for some part of documents for each topic. For each topic, the average number of judged documents is 1250 (with average vocabulary size 48810.91), and the average number of relevant documents is 74.07.

*5.1.1 Experiment setup.* Documents in TREC disks 4 & 5 are tokenized and stemmed using Snowball stemmer in NLTK toolkit [16]. After that, stop-words and punctuation are filtered for each document. As for the topic, each topic includes three parts: "title", "description" and "narrative". We preprocess the raw topic data in the same way as the documents. The average number of tokens is 2.6 in "title", 8.6 in "description" and 25.7 in "narrative". We attain the vocabulary from the judged documents for each topic, which greatly reduces the size of vocabulary for our experiments. For each document, we calculate term weight in three weighting schemes (VSM, TFIDF and BM25) and use them to initialize the query and each document vector respectively, which is indicated in Tables 3 and 4. In practice, we initialize the query using tokens in "title" and we constrain query expansion among tokens existing in "title" and "description" in each round of the game. We assume that words in "description" are the candidates for query expansion and our model is to learn the weights of these words as the query to improve the retrieval performance. In order to test the performance of our model, we apply a 3:1 random splitting on the documents in each topic where the ratio of relevance and non-relevance documents is identical in test and train sets. In relevance feedback, we assume that after each iteration the relevance signal is simulated by virtue users on the set of documents. While in pseudo relevance feedback, the top-10 documents are regarded as relevant documents in our experiment. It is reasonable to use all documents as feedback as the average number of other documents in the training set is 939.

The learning rate is 0.1 for query reformulation and 1 for retrieval model in both relevance feedback and pseudo relevance feedback. We also tried other values of learning rate and found that small learning rate for query reformulation could avoid fast query drift. If the average update on the weights of query and retrieval model falls below a threshold (we set $10^{-7}$ in the experiment), it is deemed to reach an equilibrium state. The maximum number of iteration is 2000 as most cases converge within 2000 iterations.

*5.1.2 Results and discussion.* We provide NDCG, MAP, MRR and Precision along with the standard deviation over the whole dataset. The results are displayed in Tables 3 and 4. The utility on the test set during the training process is also shown in Fig. 1. For relevance feedback, *Conv-Q* shows better performance than *naive* and *Rocchio*, which indicates that a good representation of query can greatly boost performance of text retrieval. *Conv-M* presents worse performance on the test set, although the utility of retrieval model increases during the training process. The reason might be the narrow explore space for the retrieval model as the model is a linear model with only four parameters to train. Our focus is to investigate the performance of cooperation between the query and retrieval model. More complicated models can be our future work. In addition, significant improvement can be seen in *Equil-Q&M*. It is due to the fact that the model could make use of relevance feedback to do query reformulation and also improve the retrieval model. Finally, the overall model can learn a better representation of the query and a better retrieval model to discriminate relevant and non-relevant documents. The best performance of *Equil-Q&M* shows that the query representation and the retrieval model can be (implicitly) coordinated in our game framework to finally achieve better results.

For the test on pseudo relevance feedback, we notice that *Conv-Q* has worse performance than the baseline. It is intuitively understandable because the update of query entirely relies on the top-$k$ retrieved documents from the model. The top-$k$ documents may contain much noise for the query reformulation. When there are only fewer relevant documents for a topic ($k = 10$) for instance, it is difficult for *Conv-Q* to capture the term distribution in relevant and non-relevant documents. But even under this pseudo assumption, *Equil-Q&M*, with the help of updating retrieval model, is able to overcome the situation where the query gets a worse representation, and obtains better results than both *Conv-Q* and *Conv-M*. This is consistent with our observation in our toy example in Table 2b where in the right down corner, iterative query expansion with pseudo relevance feedback would result in query drifts given the high (wrong) utility value of the query reformulation alone. However, strategically coupled with the retrieval model in the game, the new equilibrium learning shall prevent query drifts from the correct utility signal introduced by the retrieval model.

To understand the convergence of the learning process, we present the results of average NDCG, MAP, MRR and precision overtime after each iteration in Fig. 2. In relevance feedback, it is not difficult to notice that values of NDCG, MAP, precision and MRR on test set decreases slightly during the first several iterations, due to the worse performance of retrieval model on test set. But with the help of query reformulation, the performance becomes much better and finally exceeds *Conv-Q*. This proves that the coordination between the query and retrieval model can help to boost the overall performance of text retrieval. For pseudo relevance feedback, it shows more iterations do not help the retrieval model due to the reason discussed before. However, even when *Conv-M* gets worse results, the query reformulation can avoid the situation and help *Equil-Q&M* discover a better representation.

Table 3: Text retrieval results (relevance feedback) where the performance scores are given in the format of $mean \pm std$ and * indicates a significant improvement according to the Wilcoxon signed-rank test over *Rocchio* at $p < 0.05$.

| Algorithm | NDCG@10 | NDCG@30 | MRR |
|---|---|---|---|
| Naive (VSM) | 0.395±0.37 | 0.412±0.32 | 0.352±0.38 |
| Naive (TFIDF) | 0.511±0.37 | 0.528±0.33 | 0.478±0.41 |
| Naive (BM25) | 0.504±0.37 | 0.517±0.32 | 0.459±0.40 |
| Rocchio (VSM) | 0.407±0.37 | 0.422±0.32 | 0.367±0.39 |
| Rocchio (TFIDF) | 0.519±0.38 | 0.536±0.33 | 0.487±0.41 |
| Rocchio (BM25) | 0.518±0.37 | 0.531±0.32 | 0.474±0.40 |
| Conv-Q (VSM) | 0.527±0.34 | 0.554±0.29 | 0.475±0.39 |
| Conv-Q (TFIDF) | 0.568±0.35 | 0.571±0.30 | 0.530±0.40 |
| Conv-Q (BM25) | 0.563±0.35 | 0.573±0.30 | 0.522±0.40 |
| Conv-M | 0.463±0.38 | 0.482±0.34 | 0.431±0.41 |
| Equil-Q&M | **0.583±0.34** | **0.601*±0.29** | **0.537*±0.39** |
| Algorithm | P@10 | P@30 | MAP |
| Naive (VSM) | 0.152±0.18 | 0.134±0.15 | 0.184±0.16 |
| Naive (TFIDF) | 0.221±0.22 | 0.179±0.18 | 0.263±0.23 |
| Naive (BM25) | 0.217±0.22 | 0.178±0.17 | 0.262±0.23 |
| Rocchio (VSM) | 0.162±0.18 | 0.139±0.15 | 0.193±0.17 |
| Rocchio (TFIDF) | 0.225±0.22 | 0.186±0.18 | 0.276±0.24 |
| Rocchio (BM25) | 0.221±0.21 | 0.183±0.17 | 0.272±0.24 |
| Conv-Q (VSM) | 0.245±0.23 | 0.212±0.18 | 0.288±0.22 |
| Conv-Q (TFIDF) | 0.264±0.24 | 0.220±0.20 | 0.317±0.25 |
| Conv-Q (BM25) | 0.265±0.24 | 0.214±0.20 | 0.319±0.25 |
| Conv-M | 0.190±0.20 | 0.160±0.16 | 0.238±0.21 |
| Equil-Q&M | **0.278*±0.24** | **0.233±0.19** | **0.331*±0.25** |

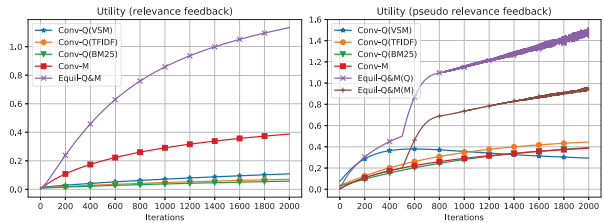

Figure 1: Utility in both cases of relevance feedback.

## 5.2 Item Recommendation

*5.2.1 User-based item recommendation.* For recommendation, we limit our experiment on the Movielens(100k) dataset [9] only, while leaving the study of scalability on a larger dataset for future work. In the user-based recommendation, we apply a 3:1 random splitting for users between the training and the test. For each user in the test set, we keep 75% item ratings as its history and initialize the rate of the other 25% items (test items) with zero. Due to the fact that there is not any ground-truth user-user similarity relevance feedback provided, only the experiment on pseudo relevance feedback is conducted. In *Rocchio* algorithm, the rate of test items is updated once according to the similarity of other user's in the memory base while other three cases follow an iterative process the same as the text retrieval. The learning rate is set to be 0.1. We also tried other values of learning rate and found that 0.1 showed the best performance. Similarly, the parameters of the retrieval model are updated 50 times within an iteration in *Conv-M* and *Equil-Q&M* as most test users converges after 50 iterations. The threshold that we consider that the query or retrieval model reaches an equilibrium state is $10^{-4}$ in this case.

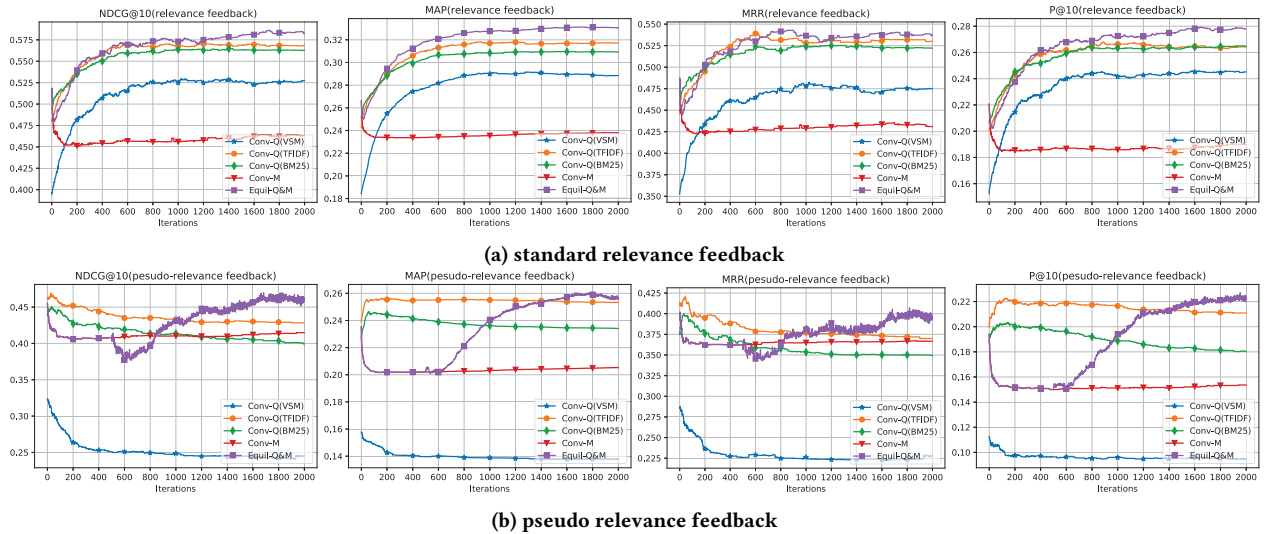

(a) standard relevance feedback

(b) pseudo relevance feedback

Figure 2: Test performance of the relevance feedback and the pesudo relevance feedback after each iteration.

Table 4: Text retrieval results (pseudo relevance feedback).

| Algorithm | NDCG@10 | NDCG@30 | MRR |
|---|---|---|---|
| Naive (VSM) | 0.323±0.38 | 0.378±0.29 | 0.287±0.36 |
| Naive (TFIDF) | 0.463±0.36 | 0.493±0.30 | **0.413±0.38** |
| Naive (BM25) | 0.439±0.35 | 0.474±0.28 | 0.375±0.36 |
| Rocchio (VSM) | 0.323±0.36 | 0.378±0.30 | 0.285±0.36 |
| Rocchio (TFIDF) | 0.460±0.36 | 0.493±0.30 | 0.410±0.38 |
| Rocchio (BM25) | 0.444±0.35 | 0.477±0.29 | 0.386±0.37 |
| Conv-Q (VSM) | 0.245±0.34 | 0.308±0.29 | 0.228±0.33 |
| Conv-Q (TFIDF) | 0.428±0.37 | 0.465±0.32 | 0.370±0.38 |
| Conv-Q (BM25) | 0.400±0.36 | 0.456±0.30 | 0.349±0.36 |
| Conv-M | 0.415±0.37 | 0.447±0.31 | 0.367±0.39 |
| Equil-Q&M | **0.469*±0.37** | **0.499±0.31** | 0.397±0.38 |

| Algorithm | P@10 | P@30 | MAP |
|---|---|---|---|
| Naive (VSM) | 0.112±0.14 | 0.100±0.12 | 0.158±0.15 |
| Naive (TFIDF) | 0.200±0.21 | 0.142±0.14 | 0.239±0.22 |
| Naive (BM25) | 0.187±0.20 | 0.137±0.13 | 0.226±0.21 |
| Rocchio (VSM) | 0.108±0.14 | 0.100±0.12 | 0.157±0.16 |
| Rocchio (TFIDF) | 0.207±0.22 | 0.145±0.14 | 0.244±0.23 |
| Rocchio (BM25) | 0.193±0.20 | 0.141±0.14 | 0.233±0.22 |
| Conv-Q (VSM) | 0.095±0.15 | 0.090±0.12 | 0.138±0.15 |
| Conv-Q (TFIDF) | 0.211±0.23 | 0.150±0.16 | 0.253±0.24 |
| Conv-Q (BM25) | 0.180±0.21 | 0.143±0.15 | 0.234±0.23 |
| Conv-M | 0.154±0.17 | 0.122±0.13 | 0.205±0.19 |
| Equil-Q&M | **0.223*±0.16** | **0.162*±0.16** | **0.257±0.23** |

Table 5: User-based recommendation results.

| Algorithm | NDCG@10 | NDCG@30 | MRR |
|---|---|---|---|
| Rocchio | 0.194±0.31 | 0.220±0.28 | 0.167±0.28 |
| Conv-Q | 0.201±0.31 | 0.234±0.28 | 0.172±0.29 |
| Conv-M | 0.199±0.32 | 0.223±0.29 | 0.170±0.30 |
| Equil-Q&M | **0.204*±0.31** | **0.237±0.28** | **0.174*±0.29** |

| Algorithm | P@10 | P@30 | MAP |
|---|---|---|---|
| Rocchio | 0.111±0.20 | 0.039±0.07 | 0.021±0.03 |
| Conv-Q | 0.113±0.19 | 0.043±0.07 | 0.024±0.03 |
| Conv-M | 0.111±0.21 | 0.040±0.07 | 0.022±0.03 |
| Equil-Q&M | **0.116±0.19** | **0.045*±0.07** | **0.025±0.03** |

The test results are shown in Table 5 and the learning curves (performance over the test users) after each iteration are shown in Fig. 3. We find that *Conv-M* did not perform well in the later stage of the iterations. *Equil-Q&M* exceeds the other cases and helps the iterative process of jointly play between the query and the retrieval model improve slightly further. Even though *Conv-M* shows worse performance in the later stage, *Equil-Q&M* can still correct the situation and exceed the performance of *Equil-Q*.

## 6 CONCLUSIONS AND FUTURE WORK

In this paper, we presented a novel equilibrium theory of information retrieval that considers a strategic game played between the query formulator and the retrieval model. We showed that the strategic game would allow us to jointly study the two correlated problems which may or may not have the same optimization objective. We provided the insights into the equilibrium learning process and discussed the resulting practical learning algorithms in the scenarios of text retrieval and collaborative filtering. The experimental results have confirmed that the equilibrium solution to (pseudo) relevance feedback would consistently outperform other competitive baseline algorithms for both text retrieval and item recommendation applications. Specifically, in the case of standard relevance feedback where both players of the game have the same utility function, our game-theoretical framework achieved superior empirical performance with a high convergence rate. Furthermore, in the case of pseudo relevance feedback, coupling the query reformulator with the retrieval model in the strategic game could help to mitigate the query drift problem.

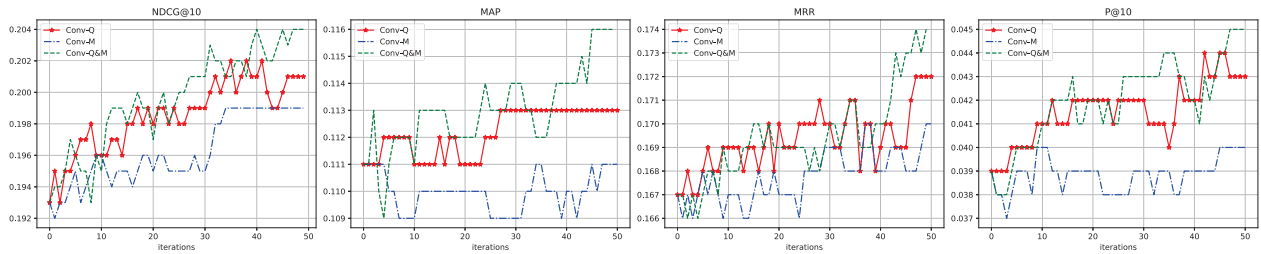

Figure 3: Test performance of the user-based item recommendation after each game iteration.

For the future work, we shall perform a deeper inquiry of utility design in the proposed normal-form IR game to further optimize the equilibrium performance of (pseudo) relevance feedback. More real-world application scenarios, including online advertising, social network recommendation, and collaborative filtering, will be investigated. We also plan to explore more efficient Nash equilibrium learning using reinforcement learning techniques [22], and extend the normal-form game to extensive-form game to deal with the dynamical nature of information retrieval [31].